\documentclass[
 reprint,
 superscriptaddress,
 amsmath,amssymb,
 aps,
 pra,
 10pt
]{revtex4-2}

\usepackage{graphicx}
\graphicspath{{figures/}}
\usepackage{dcolumn}
\usepackage{bm}
\usepackage{bbold}
\usepackage{varioref}
\usepackage[
colorlinks=true,
citecolor=blue,
urlcolor=blue,
linkcolor=blue
]{hyperref}
\usepackage[capitalise]{cleveref}
\usepackage[acronym]{glossaries}
\usepackage{braket}
\usepackage{booktabs}

\usepackage{amsthm} 
\usepackage{amsmath}
\usepackage{mathtools}

\usepackage[caption=false]{subfig}
\usepackage{xcolor}

\newcommand{\affiliationibm}{IBM Research Europe -- Zurich, S\"aumerstrasse 4, 8803 R\"uschlikon, Switzerland}
\newcommand{\affiliationepfl}{Institute of Physics, \'Ecole Polytechnique F\'ed\'erale de Lausanne (EPFL), CH-1015 Lausanne, Switzerland}
\newcommand{\affiliationeth}{Institute of Physics, \'Eidgen\"ossische Technische Hochschule Zürich (ETHZ), CH-8092 Zurich, Switzerland}
\newcommand{\affiliationOxford}{
Department of Chemistry, Oxford University, Chemistry Research Laboratory, Oxford, United Kingdom}
\newcommand{\affiliationManchester}{Department of Chemistry, The University of Manchester, Oxford Road, Manchester, United Kingdom}
\newcommand{\affiliationregen}{Institute of Experimental and Applied Physics and Halle-Berlin-Regensburg Cluster of Excellence CCE, University of Regensburg, Regensburg, Germany}

\begin{document}

\title{Exploring pathways towards quantum advantage in quantum chemistry: the case of a molecule with half-M\"obius topology}

\author{Samuele Piccinelli}
\thanks{These authors contributed equally to this work.}
\affiliation{\affiliationibm}
\affiliation{\affiliationepfl}
\author{Stefano Barison}
\thanks{These authors contributed equally to this work.}
\affiliation{\affiliationibm}
\affiliation{\affiliationeth}
\author{Alberto Baiardi}
\affiliation{\affiliationibm}
\author{Francesco Tacchino}
\affiliation{\affiliationibm}
\author{Jascha Repp}
\affiliation{\affiliationregen}
\author{Igor Ron\v{c}evi\'c}
\affiliation{\affiliationManchester}
\affiliation{\affiliationOxford}
\author{Florian Albrecht}
\affiliation{\affiliationibm}
\author{Harry L. Anderson}
\affiliation{\affiliationOxford}
\author{Leo Gross}
\affiliation{\affiliationibm}
\author{Alessandro Curioni}
\affiliation{\affiliationibm}
\author{Ivano Tavernelli}
\affiliation{\affiliationibm}
\email{ita@zurich.ibm.com}

\begin{abstract}
We report quantum chemistry calculations performed on superconducting quantum processors for a molecule exhibiting the half-M\"obius electronic topology originally introduced by Ron\v{c}evi\'c \textit{et al.}~\cite{Roncevic2026}. 
Using SqDRIFT, a randomized sample-based Krylov quantum diagonalization algorithm, we achieve reliable quantum simulations on active spaces corresponding to $36$ orbitals ($72$ qubits) and extend previous studies up to $50$ orbitals ($100$ qubits). We demonstrate that a systematic increase of active space sizes, which has a concrete impact on the accuracy of the electronic structure description, is achievable with state-of-the-art quantum processors, thus offering a promising path towards practically relevant quantum‑assisted electronic‑structure calculations.
\end{abstract}

\maketitle

\section{Introduction}

Quantum information processing offers a fundamentally new approach to scientific computing. Among the most compelling applications of this paradigm is quantum chemistry~\cite{Cao2019,McArdle2020,Motta2022,Miessen2022}, whose central challenge consists of solving the electronic Schr\"odinger equation within a chosen basis set. Indeed, the dimension of the many-electron Hilbert space grows exponentially with the number of spin orbitals, rendering exact diagonalization, i.e., full configuration interaction (FCI), intractable on classical computers beyond approximately $40$ spin orbitals, corresponding to Hilbert spaces exceeding $\sim10^{11}$ determinants~\cite{Gao2024,Shayit2025}. While lower--accuracy approximate solutions can generally be obtained with heuristic approaches, the exponential barrier has long prevented the study of large, strongly correlated systems. In this domain, quantum computers may offer a decisive advantage. Already in the current pre-fault tolerance era -- where quantum hardware composed of hundreds of qubits are becoming available, on which errors can be dealt through error-mitigation schemes~\cite{Kim2023,Berg2023,Fischer2026} -- it is therefore of particular interest to deploy quantum algorithms for chemistry at scale. This involves combining state-of-the-art quantum processors and classical high-performance resources, with the aim of establishing evidence for promising advantage candidates and exploring regimes that come progressively closer to the limits of classical methods.\\
\begin{figure}
    \centering
    \includegraphics{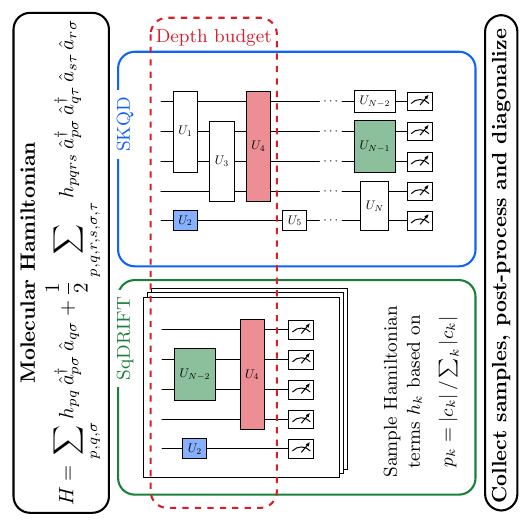}
    \caption{\textbf{The SqDRIFT algorithm.} The molecular Hamiltonian is mapped to qubit operators. Its terms $h_k$ generate time-evolution unitaries $U_k$. Circuits are executed within a fixed depth budget (dashed region). In SKQD, deterministic sequences of $U_k$ are applied, while in SqDRIFT the unitaries are sampled according to their relative importance. Measurement samples are then collected and post-processed to obtain the final energies.}
    \label{fig:sqdrift-scheme}
\end{figure}
In a recent publication~\cite{Roncevic2026}, we applied a novel sample-based quantum algorithm~\cite{RobledoMoreno2025}, SqDRIFT~\cite{Piccinelli2026}, to support the characterization of the topological electronic structure of a half-M\"obius molecule. Conceptually, SqDRIFT shares similarities with classical selected configuration interaction methods~\cite{Holmes2016} in that it constructs a relevant subspace of determinants. However, the sampling mechanism is fundamentally different. Indeed, classical methods rely on heuristic importance criteria to iteratively select the relevant determinants to be included in the subspace, whereas SqDRIFT leverages quantum time evolution itself to probe the structure of the Hamiltonian in a Krylov-subspace fashion. The resulting distribution of sampled configurations is therefore generated by the physical dynamics encoded in the Hamiltonian, offering a qualitatively distinct approach to importance sampling.
Our quantum calculations, realized on superconducting quantum processors, were performed in active spaces comprising $36$ orbitals encoded in $72$ qubits, and delivered accurate estimates of ground-state energies and electronic properties, including the experimentally measured Dyson orbitals. The results showed excellent agreement with the best available approximate classical approaches, demonstrating that sample-based quantum diagonalization can achieve chemically meaningful accuracy for large molecular systems.

In this work, we further extend our investigations to even larger active spaces, scaling the quantum simulations up to $50$ orbitals encoded in $100$ qubits. This moves the study into a regime that becomes increasingly demanding for classical post-Hartree–Fock methods and entirely inaccessible to exact multi-reference treatments. Importantly, our results reveal a non-negligible additional contribution to the total energy, highlighting the role of orbitals that were not included in the original $72$-qubit active space and its corresponding classical benchmarks. These findings show that a systematic expansion of active spaces is becoming feasible with today's quantum computers. Overall, they outline a blueprint for leveraging quantum resources in cutting‑edge chemistry research workflows.

\section{The quantum algorithm: SqDRIFT}

SqDRIFT builds upon the framework of Sample-based Krylov Quantum Diagonalization (SKQD)~\cite{Yu2025}.
In the spirit of Krylov quantum diagonalization, for a given Hamiltonian $H$, SKQD constructs a subspace generated by successive applications of the time-evolution operator to an initial reference state $\ket{\Psi_0}$,
\begin{equation}
  \mathcal{K} = \left\{ \ket{\Psi_0}, \mathrm{e}^{-\mathrm{i} H t} \ket{\Psi_0},
          \mathrm{e}^{-2 \mathrm{i} H t} \ket{\Psi_0}, \ldots \right\} \,.
  \label{eq:Krylov}
\end{equation}
Rather than diagonalizing $H$ in the subspace defined by $\mathcal{K}$, which requires deep circuits, SKQD samples bitstrings from time-evolved states (each bitstring corresponding to a Slater determinant, where a $0$ indicates an empty orbital, and a $1$ an occupied one).
These bitstrings define a subspace in which the Hamiltonian is then projected and diagonalized classically.

The applicability of SKQD to quantum-chemical problems is hindered by the fact that the time-evolution operator $\mathrm{e}^{-\mathrm{i} H t}$ is encoded by deep circuits even for the simplest systems.
SqDRIFT~\cite{Piccinelli2026} bypasses this limit and enables practical quantum chemistry applications by leveraging the qDRIFT randomized compilation protocol~\cite{Campbell2019}.
Given an Hamiltonian $H$ expressed as
\begin{equation}
  H = \sum_k c_k h_k\,,
  \label{eq:HamiltonianDecomposition}
\end{equation}
(where $h_k$ stands for a string of Pauli operators~\cite{Barkoutsos2018}), qDRIFT approximates the quantum channel defined by the time-evolution operator $\mathrm{e}^{-\mathrm{i} H t}$ as
\begin{equation}
  \mathcal{E}_\text{qDRIFT}[\rho] = \sum_{\bm{k}} p_{\bm{k}} V_{\bm{k}} \rho V_{\bm{k}}^\dagger\,.
  \label{eq:Average_qDRIFT_Channel}
\end{equation}
Here each randomized evolution sequence 
\begin{equation}
  V_{\bm{k}} = \prod_{j=1}^N \mathrm{e}^{- \mathrm{i} h_{k_j} t \lambda / N} \,
  \label{eq:qDRIFT_Unitary}
\end{equation}
is constructed by sampling $N$ terms $h_k$ from \cref{eq:HamiltonianDecomposition} with probability $p_k$ proportional to their coefficient $\left| c_k \right|$ and $\lambda = \sum_k \left| c_k \right|$ (see also \cref{fig:sqdrift-scheme}). qDRIFT yields an improvement over conventional Trotter formulas for sparse Hamiltonians, whose norm does not depend on (or changes very slowly with) the system size, as is the case for quantum-chemistry Hamiltonians.

SqDRIFT inherits the formal convergence guarantees of SKQD~\cite{Piccinelli2026}: if the ground state is sufficiently concentrated, i.e. it has support on a subset of basis states that grows at most polynomially with the system size, and the initial state has non-negligible overlap with it, sufficiently deep SqDRIFT circuits are guaranteed to generate bitstrings where the ground-state wave function has large support.
In practice, SqDRIFT trades circuit depth for sampling overhead (see also \cref{fig:sqdrift-scheme}): by replacing a single, deep Trotter circuit with many approximate, shallower ones, it enables tackling complex quantum-chemical Hamiltonians.
Notably, as in SKQD, the quantum circuits used in SqDRIFT are non-parametric and therefore do not require iterative optimization, thereby avoiding the substantial classical and quantum overhead typically associated with variational procedures.
\begin{figure} 
    \includegraphics[width=0.50\linewidth, angle=-90]{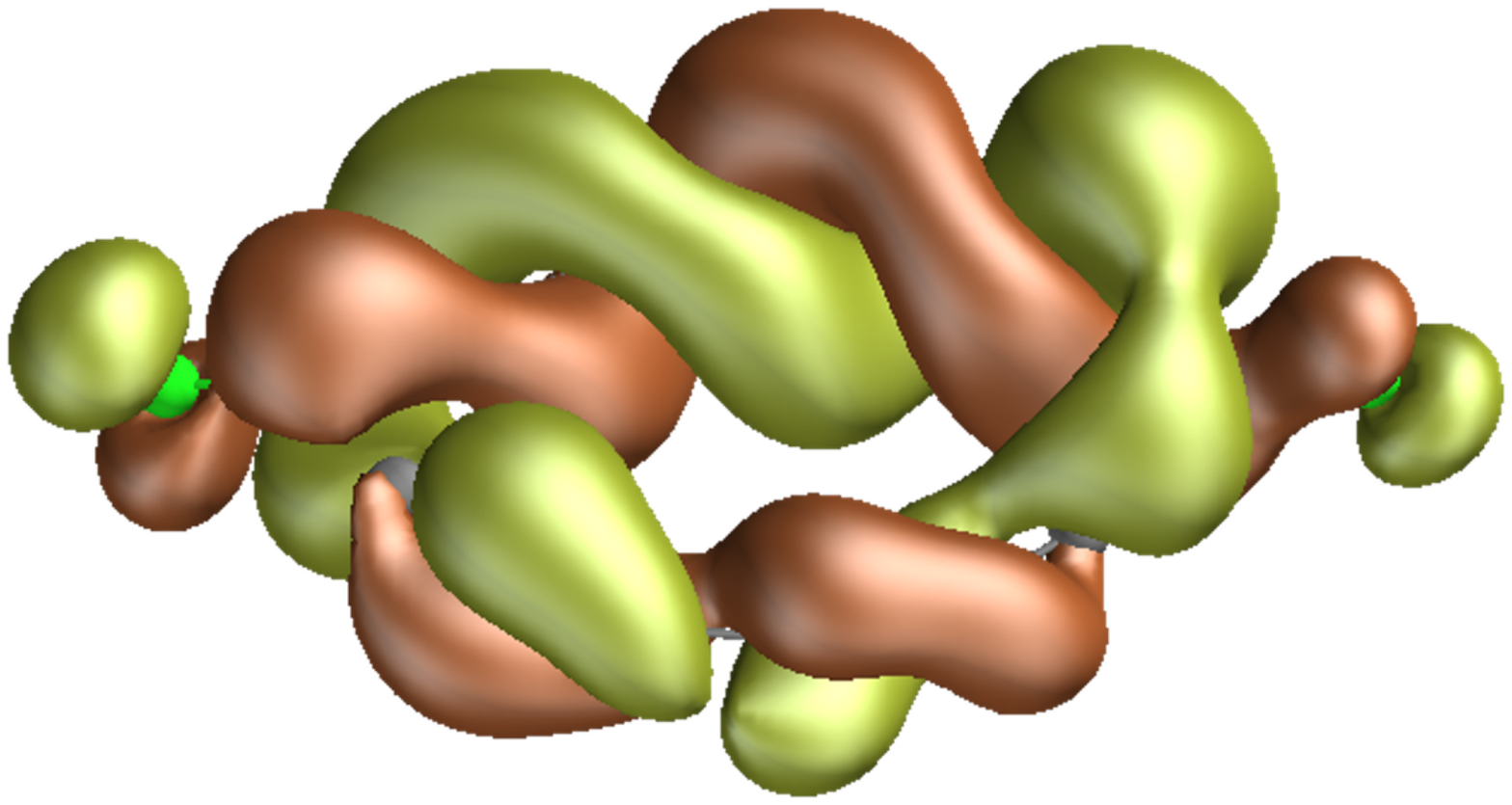}
    \caption{\textbf{Dyson orbital.} Isosurface of the Dyson orbital computed with the SqDRIFT algorithm with $72$ qubits, corresponding to an active space of $32$ electrons in $36$ orbitals.}
    \label{fig:moebius-dyson}
\end{figure}
\section{The system: The Half-M\"obius Molecule}

In this paper, we deepen our analysis of the electronic structure properties of a topological stereoisomer of the molecule $\mathrm{C_{13}Cl_2}$ in its (singlet) ground state, synthesized and characterized in Ref.~\cite{Roncevic2026}.
The $\pi$-orbital basis of this ring molecule twists by $90^\circ$ upon one circumnavigation of the carbon ring, corresponding to a generalized M\"obius-Listing topology.
This topology, which was named half-M\"obius, differs fundamentally from both trivial H\"uckel systems (no twist) and conventional M\"obius systems ($180^\circ$ twist).
In a half-M\"obius topology the $\pi$-basis changes sign only after two full circulations and becomes periodic after four. 
A quasiparticle on a ring with this boundary condition could be interpreted as carrying a Berry phase of $\pi/2$. 

The half-M\"obius topology represents a rare example of a molecular system whose electronic structure is directly shaped by nontrivial topology, making it an ideal testbed for advanced quantum-chemical simulation techniques.
In our previous work~\cite{Roncevic2026}, we simulated the Dyson orbital with SqDRIFT (see \cref{fig:moebius-dyson}), a one-particle observable that has direct correspondence with STM imaging.
We showcased how SqDRIFT can be used to reliably predict the non-trivial electronic topology of the half-M\"obius molecule.
Here we focus on the SqDRIFT calculation of the total energy of the neutral ground electronic state of the same molecule and compare it with state-of-the-art classical methods.

\section{Results}
\begin{figure}
    \centering
    \includegraphics[width=\columnwidth]{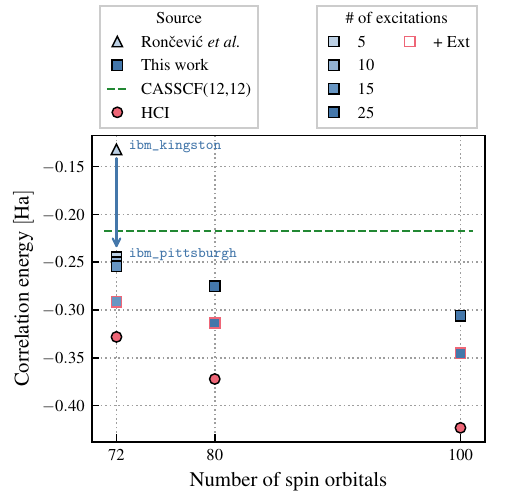}
    \caption{\textbf{Correlation energy (relative to Hartree–Fock) as a function of the active space size (number of spin orbitals/qubits) at a fixed number of sampled configurations for the diagonalization (subspace dimension $\bm{\mathrm{d}}\approx 10^7$).} Blue points correspond to the energy evaluated with the (extended (Ext)~\cite{Barison2025}, red contours) SqDRIFT algorithm as described in~\cite{Piccinelli2026}. The triangle corresponds to the best energy value obtained with the data of Ref.~\cite{Roncevic2026}. The squares are instead obtained using improved hardware and extended to a growing number of spatial orbitals. For the classical calculations, the green dashed line shows the CASSCF results for $12$ electrons in $12$ orbitals and the red circles correspond to the energies computed with HCI with an equivalent number of samples as for the SqDRIFT calculations.}
    \label{fig:moebius-convergence}
\end{figure}
We characterize the convergence behavior of SqDRIFT with the active space size.
More specifically, we executed SqDRIFT simulations on quantum hardware for a sequence of systematically increasing active spaces with $36$ (already reported in~\cite{Roncevic2026}), $40$, and $50$ orbitals, corresponding to $72$, $80$, and $100$ qubits, respectively.
The subspace dimension is fixed, in all cases, to $\mathrm{d}^2\leq 3165^2$.
Both SqDRIFT and extended SqDRIFT (ExtSqDRIFT)~\cite{Barison2025} results are reported.
The ExtSqDRIFT calculations were carried out by generating singly- and doubly-excited determinants starting from the ones obtained through SqDRIFT.

As shown in \cref{fig:moebius-convergence}, we obtain consistently lower energy estimates with $80$ and $100$ orbitals compared to the energy estimates obtained with $72$ orbitals in Ref.~\cite{Roncevic2026}, in line with the same decreasing trend observed for classical reference calculations. Furthermore, we improve on the same results published in Ref.~\cite{Roncevic2026} (blue arrow in \cref{fig:moebius-convergence}). The possibility of obtaining better energy estimates in this case is the result of a combination of better hardware quality -- the $100$-qubit calculations were performed on the latest generation of Heron r3 quantum processors, \texttt{ibm\_pittsburgh} -- and a more effective error mitigation, based on a self-consistent configuration recovery scheme with a carry-over of the most important determinants, first proposed in Ref.~\cite{Shirakawa2025}. These initial observations indicate that, even at the scales considered here, quantum computers can produce meaningful results. More importantly, the underlying quantum approach offers a fundamentally scalable route to larger systems.

We compared our results with two reference classical methods. Specifically, we considered (i) a complete active space self-consistent field (CASSCF) calculation carried out in an active space comprising $12$ electrons and $12$ orbitals -- a size that can be tackled classically but that at the same time is already near the upper limit of routinely feasible CASSCF calculations -- and (ii) a Heat-Bath Configuration Interaction (HCI) obtained on $40$ and $50$ orbitals. This corresponds to the same subspace size studied with SqDRIFT.

For $72$ orbitals, the new SqDRIFT energy estimates surpass the CASSCF reference energy. Although the energies remain higher than those returned by HCI, the convergence properties of SqDRIFT~\cite{Piccinelli2026} suggest that the accuracy can be further improved. This may be achieved by increasing the number of excitations included in each qDRIFT randomization and through continued improvements in quantum hardware, as already mentioned and demonstrated by the $72$-orbital case. Moreover, although ExtSqDRIFT is necessary to improve the energies, this step builds on the subspace identified by SqDRIFT. The need for this extension reflects current hardware limitations, while the quantum procedure already identifies the relevant sector of the Hilbert space on which the classical expansion can build. Finally, although the computed energies may exhibit some dependence on the selected subspace, a detailed investigation of this effect is deferred to future work. We do not anticipate significant changes to the conclusions, and such an analysis lies beyond the scope of the present study.

\section{Conclusions and Outlook}

The results presented here mark a meaningful step forward in the development of quantum computing methods for quantum chemistry.
By combining Krylov-based sampling with qDRIFT time evolution, SqDRIFT provides a hardware-efficient algorithm with formal convergence guarantees that can be executed on present-day quantum processors~\cite{Piccinelli2026}. Its application to a molecule exhibiting half-M\"obius electronic topology demonstrates that quantum computation can now address experimentally relevant systems characterized by a nontrivial orbital topology~\cite{Roncevic2026}.

Our scaling study indicates that enlarging the active space from $72$ to $100$ orbitals yields improved correlation energies at a size where classical exact solvers are excluded.
This improvement is achieved using the same number of quantum circuits, the same subspace dimension and comparable circuit depths -- i.e., without increasing the QPU and classical HPC resources required to obtain these energy estimates.

In summary, this work illustrates how hybrid quantum-classical workflows can begin to explore regions of the Hilbert space inaccessible to exact classical algorithms, marking a transition from proof-of-principle demonstrations towards practical quantum-enhanced electronic structure calculations.

\begin{acknowledgments}
We thank Max Rossmannek and Conrad Haupt for valuable feedback on the manuscript. This research was supported by RESQUE funded by the Swiss National Science Foundation (grant number 225229).
\end{acknowledgments}

\bibliography{biblio}

\end{document}